# Surface Segregation in Multicomponent High Entropy Alloys: Atomistic Simulations versus a Multilayer Analytical Model

Dominique Chatain[1] and Paul Wynblatt[2,*]

[1]*Aix-Marseille Univ, CNRS, CINaM, Marseille, France*
[2]*Department of Materials Science and Engineering, Carnegie Mellon University, Pittsburgh PA 15213, USA*

**Abstract**

This paper compares two approaches for investigating the near-surface composition profile that results from surface segregation in the so-called Cantor alloy, an equi-molar alloy of CoCrFeMnNi. One approach consists of atomistic computer simulations by a combination of Monte Carlo, molecular dynamics and molecular statics techniques, and the other is a nearest neighbor analytical calculation performed in the regular solution approximation with a multilayer model, developed here for the first time for a N-component system and tested for the 5-component Cantor alloy. This type of comparison is useful because a typical computer simulation requires the use of ~100 parallel processors for 2 to 3 hours, whereas a similar calculation by means of the analytical model can be performed in a few seconds on a laptop machine. The results obtained show qualitatively good agreement between the two approaches. Thus, while the results of the computer simulations are presumably more reliable, and provide an atomic scale picture, if massive computations are required, for example, in order to optimize the composition of a multicomponent alloy, then an initial screening of the composition space by the analytical model could provide a highly useful means of narrowing the regions of interest, in the same way that the CALPHAD method allows rapid investigation of phase diagrams in complex multinary systems.

---

[*] corresponding author

**Highlights**

* Surface segregation in a 5-component alloy is investigated by 2 approaches.

* The approaches consist of atomistic simulations and an analytical method.

* Results obtained by the two methods are qualitatively similar.

* The analytical approach is far more economical in computational resources.

* The latter is useful for mapping the composition dependence of multinary properties.

**Graphical abstract**

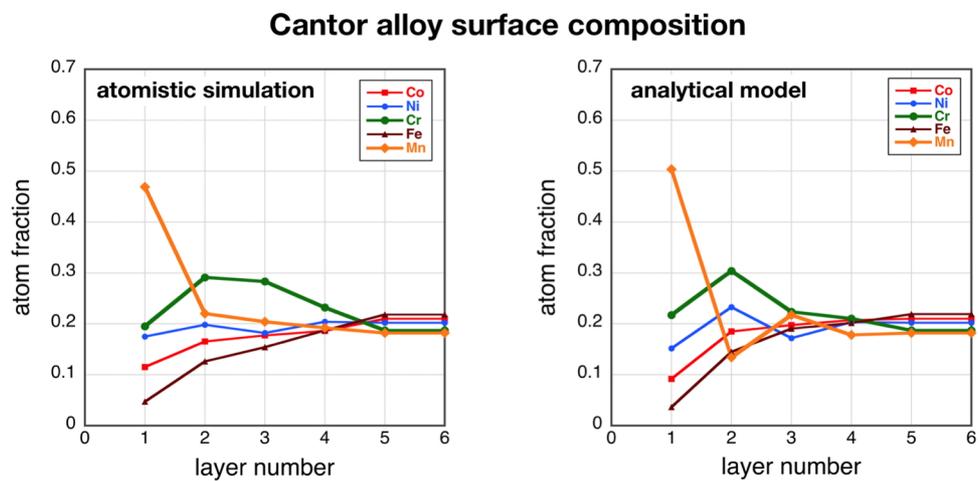

# 1. INTRODUCTION

Most of the work on interfacial segregation in multi-component alloys consisting of five or more elements has been conducted on the Cantor alloy (an equi-molar combination of Co, Cr, Fe, Mn, and Ni) [1-5]. However, this literature is quite sparse, as only three of these papers have addressed modeling issues [1,2,5], and only one has dealt with surface segregation. Some of these papers [3-5] have reported experimental measurements on grain boundaries, but the measurements were conducted at temperatures where the alloys consist of more than one phase [6], thereby complicating the interpretation of the results.

Our own previous work [1] has consisted of atomistic simulations of interfacial segregation, performed by a combination of Monte Carlo (MC), molecular dynamics (MD) and molecular statics (MS) approaches, used in conjunction with modified embedded atom method (MEAM) potentials, developed specifically for the case of systems consisting of the five Cantor alloy components [7]. A typical computer simulation performed by our approach requires the use of ~100 parallel processors for 2 to 3 hours on a computation cell consisting of 8000 atoms. In the present paper, we develop an analytical model for N-component surfaces using a simple nearest neighbor approach in the regular solution approximation, which describes the surface as a stack of atom layers, for consistency with the Gibbs adsorption isotherm. The analytical model is tested for the 5-component Cantor alloy and its underlying binaries. While the predictions of this analytical approach are presumably less reliable than those of the atomistic simulations, a computation by means of the analytical model takes no longer than a few seconds on a typical laptop. Thus, one of the issues investigated in this paper is the possibility of using the multilayer analytical model to map out general trends in multinary alloys, as a guide to subsequent massive atomistic simulations that produce presumably more reliable results.

The primary thrust of our previous work [1] was on grain boundary segregation, although some attention was also paid to surface segregation. In contrast, the focus of the present study will be on surface rather than grain boundary behavior, because the available analytical models for the description of segregation at surfaces contain fewer adjustable parameters than those for grain boundaries [8].

We also investigate whether the results of segregation in a quinary system can be inferred from results obtained on the underlying binary systems, because the inputs needed to model two-component systems are much more readily available than those suitable for a five-component system.

As mentioned in a paper that has investigated significant portions of the Co-Cr-Fe-Mn-Ni phase diagram by a combination of experiments and CALPHAD calculations [6], the CALPHAD method is appropriate for the study of high entropy alloy (HEA) phase diagrams. Similarly, this paper aims to test an easily implemented analytical approach, using the regular solution approximation, to model surface behavior, in order to establish the extent to which such an approach might be valid for studying interfacial segregation in HEAs. Finally, this model can also provide a useful tool to map the variation of properties over the whole complex composition space of multinary alloys, for comparison to high-throughput experimental data, such as that recently reported for the catalytic properties of a ternary alloy system [9].

## 2. METHODOLOGY
### 2.1. Computer simulations

Computer simulations were conducted by means of a combination of Monte Carlo (MC), molecular dynamics (MD) and molecular statics (MS) types of simulations, employing the LAMMPS code [10-12] in conjunction with second nearest neighbor (2NN) MEAM potentials [7]. The computation cells for modeling surface segregation consisted of 8000 atoms of either an equi-atomic combination of the five Cantor alloy components, or of a pair of components chosen from among the five Cantor components for the binary alloy simulations. In either case, the components were initially randomly distributed on face centered cubic (FCC) crystal sites, arranged in the shape of a rectangular solid, with periodic boundary conditions in the x- and y-directions, and terminated by two (001) surfaces normal to the z-direction.

All the simulations were performed for a temperature of 1200 K. At this temperature, the Cantor alloy adopts the FCC phase [13,14]. Simulations for the ten possible underlying binary systems were performed on alloys consisting nominally of 20 at% of one component and 80 at% of the other. The actual bulk compositions used were slightly different, as described in the section on RESULTS. The method for selecting the solute element will also be described later.

As a first step in all simulations, the computation cells were relaxed by MS, and then subjected to a certain number of cycles, each of which included MC, MD and MS modeling, in sequence. A canonical simulation was used in the MC segment of the cycle. This MC segment involved 100,000 attempted atom swaps for the case of the 5-component alloy, and 40,000 attempted swaps for the solute element in the case of binary alloys. MC was followed by MD, which consisted of two stages, a gradual increase of the temperature to 1200 K over about 12 picoseconds, and an equilibration at 1200 K for an additional 12 picoseconds. Finally, the computation cell was relaxed by MS at 0 K to remove thermal noise. This type of MC-MD-MS cycle was repeated 4 to 6 times in order to generate reliable near-surface composition profile statistics. The results showed that the near-surface composition excursion that resulted from surface segregation only extended over the first 4 atom-layers, so that the regions beyond the $4^{th}$ atom-layer were assigned the average bulk composition.

### 2.2 Analytical model

The analytical model is derived from one developed previously for liquid binary alloys, in which a FCC-like structure was assumed [15]. It is a multilayer model in which the atoms occupy the sites of a FCC lattice. The solid solution is considered to be regular, although a subregular version of the model has also been published [16]. It has been shown that this model obeys the Gibbs adsorption equation. This is because the interface consists of several atom layers rather than being confined to a single layer, as in some other models of surface or grain boundary segregation [5,17]). As a result, it is possible to calculate the equilibrium near-surface compositions that correspond to minimum surface energy. The relation used to compute the surface energy takes into account contributions related to the atom-layer chemistry as well as the elastic energy due to atom size differences [18,19], although the latter contribution turns out to be negligible in the Cantor alloy as explained below.

In the case of a N-component alloy, the energy, $\gamma$, of the segregated surface may be written as:

$$\begin{aligned}
\gamma\Omega &= \sum_{n=1}^{N}(y_n^1 \gamma_n^0)\,\Omega + RT \sum_{i=1}^{L}\sum_{n=1}^{N} y_n^i \ln\left(\frac{y_n^i}{x_n}\right) \\
&+ z_v \sum_{1\leq n<m\leq N} \omega_{nm}(x_n x_m - x_n y_m^1 - x_m y_n^1) \\
&+ z_v \sum_{i=1}^{L}\sum_{1\leq n<m\leq N} \omega_{nm}\big[(y_m^i - x_m)(y_n^{i+1} - x_n) + (y_n^i - x_n)(y_m^{i+1} - x_m)\big] \\
&+ z_l \sum_{i=1}^{L}\sum_{1\leq n<m\leq N} \omega_{nm}(y_n^i - x_n)(y_m^i - x_m) \\
&- E(N)^{el}
\end{aligned} \quad (1)$$

where the indices n and m refer to the N components; the index i, with values from 1 to L counts atom surface layers, with layer 1 being the outermost surface layer, and L=4 for consistency with the simulation results; $\Omega$ is the surface area per mole of atoms of the FCC{001} surface, which has been computed to be 39440 m$^2$/mol from the average diameter (0.508 nm [1]) of an atom of the 5-component Cantor alloy; the $\gamma_n^0$ are the surface energies of the N pure components; $x_n$ and $x_m$ are the bulk atom fractions of the components; $y_n^i$ and $y_m^i$ are the atom fractions of the components in the ith surface layer; $\omega_{nm}$ is the regular solution parameter for the binary nm alloy, (i.e., it is equal to the $^0L$ Redlich-Kister parameter [20]); $z_l$ and $z_v$ are numbers of nearest neighbors of an atom, where $z_l$ is the number of neighbors that lie in the same atom plane as the atom of interest, and $z_v$ is half the number of out-of-plane neighbors, ($z_l$ and $z_v$ define the orientation of the surface plane, e.g. for values of 4 and 4, respectively, they define a FCC {001} surface; and $E(N)^{el}$ is a measure of the change in elastic strain energy when atoms in the bulk are exchanged with atoms located in the near-surface segregated region.

To facilitate the interpretation of this equation, we present here the expression for the case of a non-equimolar two-component alloy consisting of A-type solvent atoms and B-type solute atoms:

$$\gamma \Omega = y_A^1 \gamma_A^0 \Omega + y_B^1 \gamma_B^0 \Omega + RT \sum_{i=1}^{L} \left[ y_A^i \ln\left(\frac{y_A^i}{x_A}\right) + y_B^i \ln\left(\frac{y_B^i}{x_B}\right) \right]$$

$$+ z_v \, \omega_{AB} (x_A x_B - x_A y_B^1 - x_B y_A^1)$$

$$+ z_v \, \omega_{AB} \sum_{i=1}^{L} [(y_B^i - x_B)(y_A^{i+1} - x_A) + (y_A^i - x_A)(y_B^{i+1} - x_B)]$$

$$+ z_l \, \omega_{AB} \sum_{i=1}^{L} (y_A^i - x_A)(y_B^i - x_B)$$

$$- E(2)^{el} \qquad (2)$$

The elastic strain energy term, in the case of a binary alloy, can be written out explicitly as [18,19]:

$$E(2)^{el} = \frac{24\pi K_B G_A r_A r_B (r_A - r_B)^2}{3 K_B r_B + 4 G_A r_A} \sum_{i=1}^{L} [(y_B^i - x_B) F^i] \qquad (3)$$

where the leading fraction is an expression due to Friedel [21] for the strain energy of a solute atom in the bulk, $K_B$ is the bulk modulus of the solute, $G_A$ is the shear modulus of the solvent, $r_A$ and $r_B$ are the atomic radii of the pure solvent and solute atoms, respectively, and the function $F^i$ is an empirical expression that accounts for the plane into which the solute atom is exchanged [18,19].

It should be mentioned that the elastic strain energy terms for both the binary and multinary alloys, i.e. the terms $E(N)^{el}$ in Eq. 1 and $E(2)^{el}$ in Eq. 2, have been omitted in the analytical calculations of both the binary and quinary systems. This is because no suitable expression for the elastic contribution has been developed yet for a multinary system, but also, as can be seen from Eq. 3, because the elastic energy contribution depends on the square of the size difference between atoms of the solute and solvent: $(r_A-r_B)^2$. For both the Cantor alloy and all the underlying binary systems, $(r_A-r_B)$ in the FCC structure, for all pairs of the five elements, is in the range of ± 0.003 nm [1]. Thus, the contribution of strain energy is negligible. This is why it has been possible to ignore the elastic term in the present analytical evaluations of segregation in both the quinary and binary alloys.

Returning to Eqs. 1 and 2, it is useful to identify the different terms that make up the energy of the segregated surface. The first term represents the effects of the surface energies of the pure components; the second term containing the natural logarithms of compositions arises from entropy contributions; the third, fourth and fifth terms, all of which contain the regular solution parameter $\omega_{nm}$ or $\omega_{AB}$ as a factor, represent alloy interactions that introduce phase diagram features into the surface energy, and the sixth and final term is the elastic strain energy.

The solution method we have used for Eqs. 1 and 2, consists of minimizing the surface energy with respect to the near-surface layer compositions, using a Monte Carlo method, which takes into account the constraint that the sum of the atom fraction variables in each layer is equal to unity. Thus, there are L variables for the case of a binary alloy (one independent composition for each of the L atom layers), and 4L in a quinary (and (N-1)L in a N-multinary). Such calculations yield the minimum energy of the segregated surface as well as the near-surface composition profile of each adsorbed species.

In order to provide a proper comparison between the results of simulations and those of the analytical model, we have evaluated certain constants used in the analytical model by means of the MEAM potentials used in the simulations. This includes values of the surface energies of the five pure elements of the Cantor alloy in the FCC structure [1], as shown in Table 1, as well as the regular solution parameters of the ten underlying binary alloys, shown in Table 2, also for the alloys in the FCC crystal structure.

**Table 1.**
Computed energies of the (100) surface (mJ/m$^2$) of the Cantor alloy elements in the FCC structure [1].

| Element | Surface energy |
|---------|----------------|
| Co      | 2117           |
| Ni      | 1938           |
| Cr      | 2122           |
| Fe      | 2371           |
| Mn      | 1600           |

**Table 2.**
Computed regular solution constant (kJ/mole), at infinite dilution, for the binary alloys between Cantor elements, in the FCC structure.

|     | Co     | Ni     | Cr      | Fe     |
|-----|--------|--------|---------|--------|
| Ni  | -0.579 |        |         |        |
| Cr  | 1.389  | 0.008  |         |        |
| Fe  | -2.755 | -1.441 | 2.5436  |        |
| Mn  | -2.217 | -3.683 | -1.7303 | -0.382 |

## 3. RESULTS

The results of simulations on the Cantor alloy at a temperature of 1200 K are presented in Fig. 1a. The figure shows a near-surface composition profile in the form of a plot of the variation of the atom fractions of the five components as a function of the (002) atom layers, numbered from the outermost surface atom layer, averaged over all MC-MD-MS cycles. As mentioned above, the segregated near-surface composition excursion only extends over the first 4 atom-layers, so that atom layers beyond the 4th have been assigned the average bulk composition. It should be noted that these average bulk compositions are somewhat different from the initial equi-atomic compositions. This is a result of the canonic MC approach used in the simulations, which conserves the total numbers of atoms of each type in the computation cell. Thus, the species that segregate to the surface undergo depletion of their bulk concentrations, and vice versa, although in no case is the bulk composition changed by more than 2 at%. Figure 1b shows the composition profile for the Cantor alloy obtained by the analytical model. Since the bulk compositions are an input parameter in the analytical model, we have used the bulk compositions obtained in the simulations to compute the near-surface composition profiles shown in Fig. 1b.

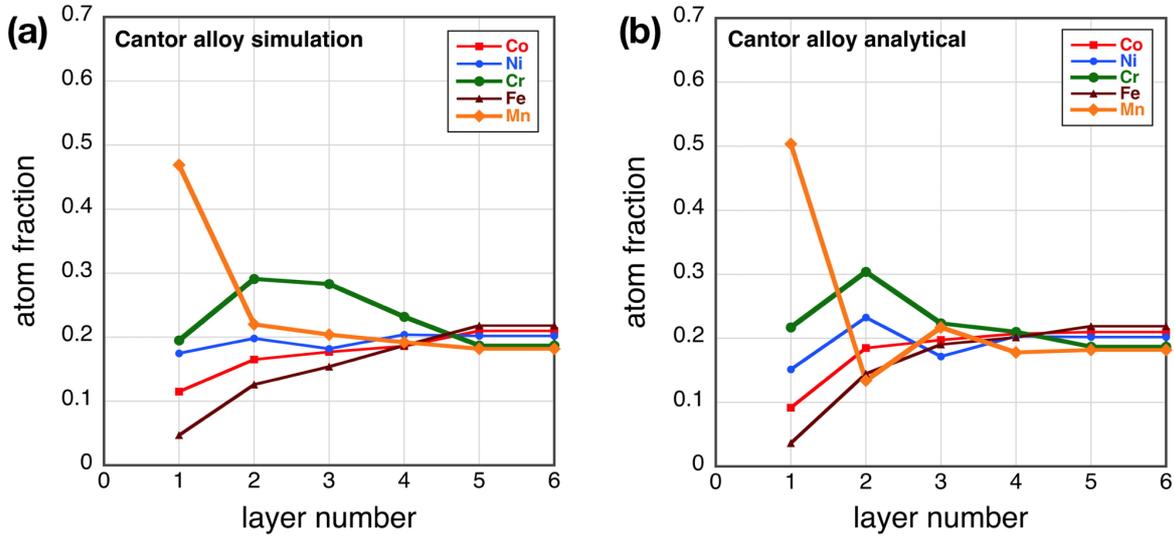

**Fig. 1**. Near-surface composition profiles of the Cantor alloy after equilibration at 1200 K. (a) Averaged atom fraction composition profile obtained by computer simulation, calculated at each (002) plane. (b) Corresponding near-surface composition profile obtained for the Cantor alloy by the analytical model. (See text).

In the case of the binary alloy simulations, we have selected as solute the element which segregated more strongly in the computer simulation of the quinary alloy, as shown in Fig. 1a. Thus, for example, for the Co-Ni pair, Ni was chosen to be the solute, whereas for the Ni-Mn pair, Mn was taken to be the solute. In addition, the initial solute concentration of the binary computation cells was set to 20 at%, for suitable comparison with the quinary alloy. In the event that the bulk composition of the solute changed by more than 1 at% due to depletion of the segregant, we made use of the bulk composition obtained in simulations as the input to the analytical model.

Although we have attempted to perform the simulations so as to obtain results on the binaries with the FCC structure, for comparison with the behavior of the FCC quinary, that turned out not to be possible. For example, in the binary Cr-Mn system, the phase diagram does not display a FCC phase. As a result, if the atoms of our selected Cr-20 at% Mn alloy are initially placed on a FCC lattice, and the simulation is performed at 1200 K, the structure is unstable and transforms to BCC, as expected from the phase diagram. So, we do not present the results of simulations for the Cr-Mn case. Also, in the Fe-Cr system, according to the phase diagram, the Fe-20 at% Cr alloy is BCC at 1200 K. Although the phase diagram indicates that the Fe-10 at% Cr alloy would be FCC at 1200 K, simulations using that composition also produced a BCC structure. The problem is a defect of the MEAM Cantor alloy potentials [7], for which even pure iron does not undergo a transformation from BCC to FCC around 1180 K, as it should. However, this defect does not invalidate the set of MEAM potentials, since they predict the correct phases in the Cantor alloy, for which the set of potentials was developed. Thus, we also omit any simulation results on the binary Fe-Cr system in this paper.

Since Mn is the strongest segregating element in the quinary simulation results of Fig. 1a, Mn was taken to be the solute in all Mn-containing binaries. Figures 2a, 2c, and 2e show the results of all the computer simulations of Mn-containing binaries that display the FCC structure at 1200 K. Figures 2b, 2d and 2f show the corresponding results obtained by the analytical model. The Cr-Mn system has been omitted, as mentioned above, because of the absence of the FCC structure in that system. Note also, that because of the relatively strong segregation of Mn to the Fe-Mn surface, the bulk concentration of Mn has been depleted to 17.4 at%.

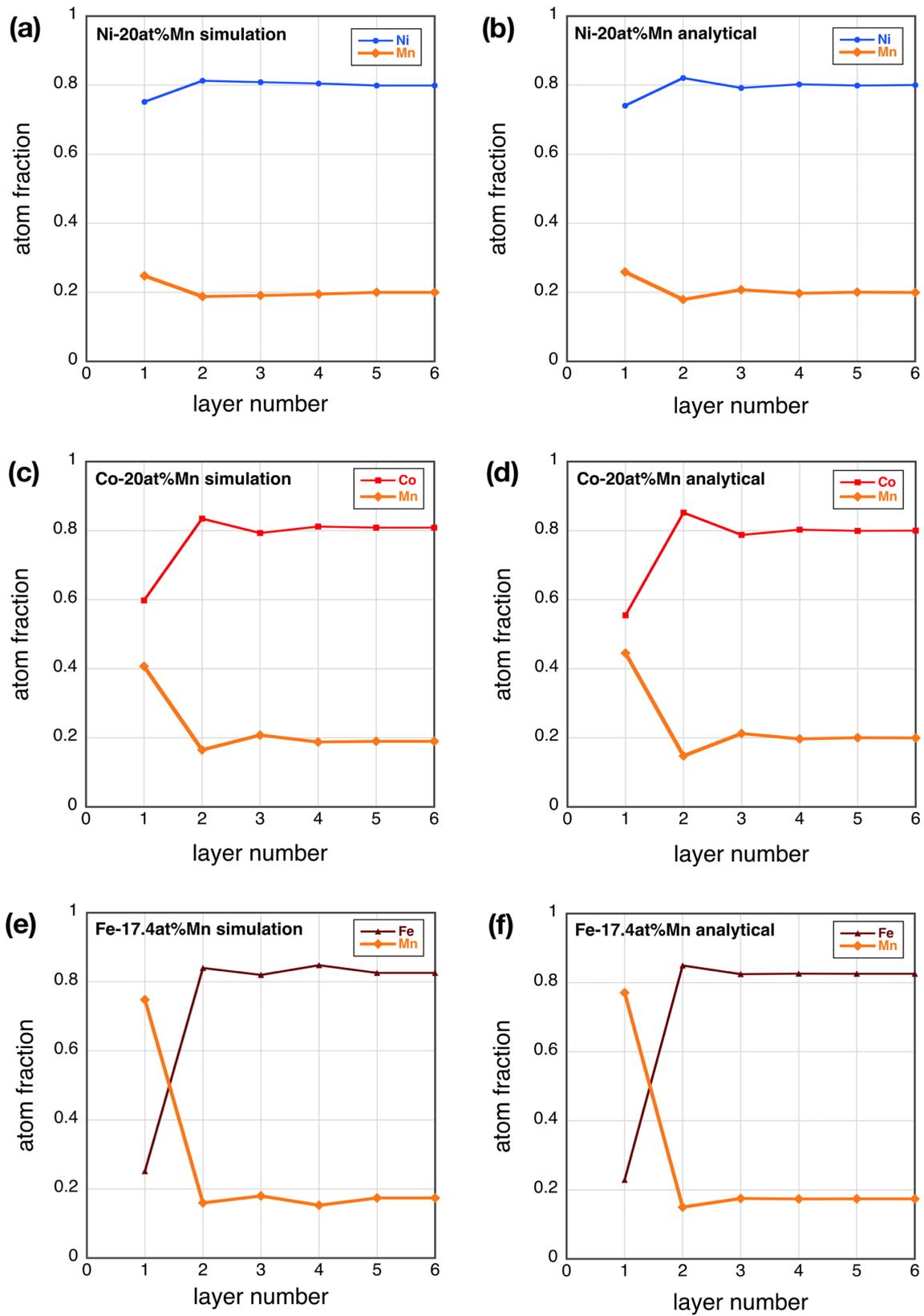

Fig. 2. Near-surface composition profiles of systems with a Mn solute, at 1200 K. The left column gives the results of simulations, and the right column gives those computed by the analytical model. (a) and (b) Ni-20 at% Mn, (c) and (d) Co-20 at% Mn, (e) and (f) Fe-17.4 at% Mn.

Figures 3a and 3c show the results of simulations for the Ni-Cr and Co-Cr systems (with Cr as the solute) and Figs. 3b and 3d provide the corresponding results by the analytical model. Note that no Fe-Cr results are shown because of the inability, mentioned earlier, of obtaining simulation results for that system in the FCC structure.

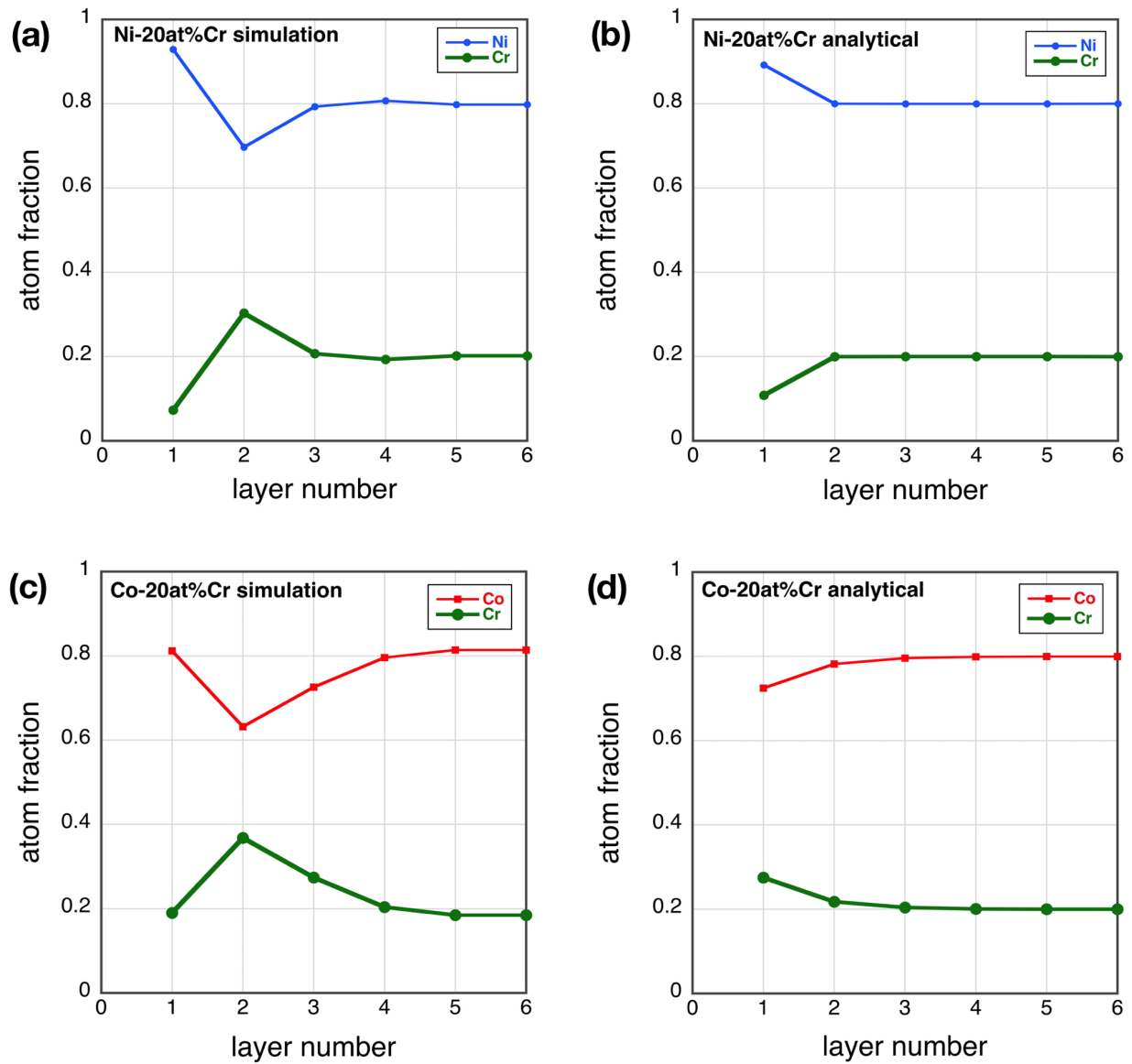

**Fig. 3**. Near-surface composition profiles of systems with a Cr solute, at 1200 K. The left column gives the results of simulations, and right column those computed by the analytical model. (a) and (b) Ni-20 at% Cr, (c) and (d) Co-20 at% Cr.

The results of systems with Ni as the solute, Co-Ni and Fe-Ni are shown in Figs. 4a to 4d, and the Fe-Co system with Co as solute is shown in Figs. 4e and 4f. Note that the Ni segregates strongly enough in the Fe-Ni system to require re-adjustment of the bulk composition.

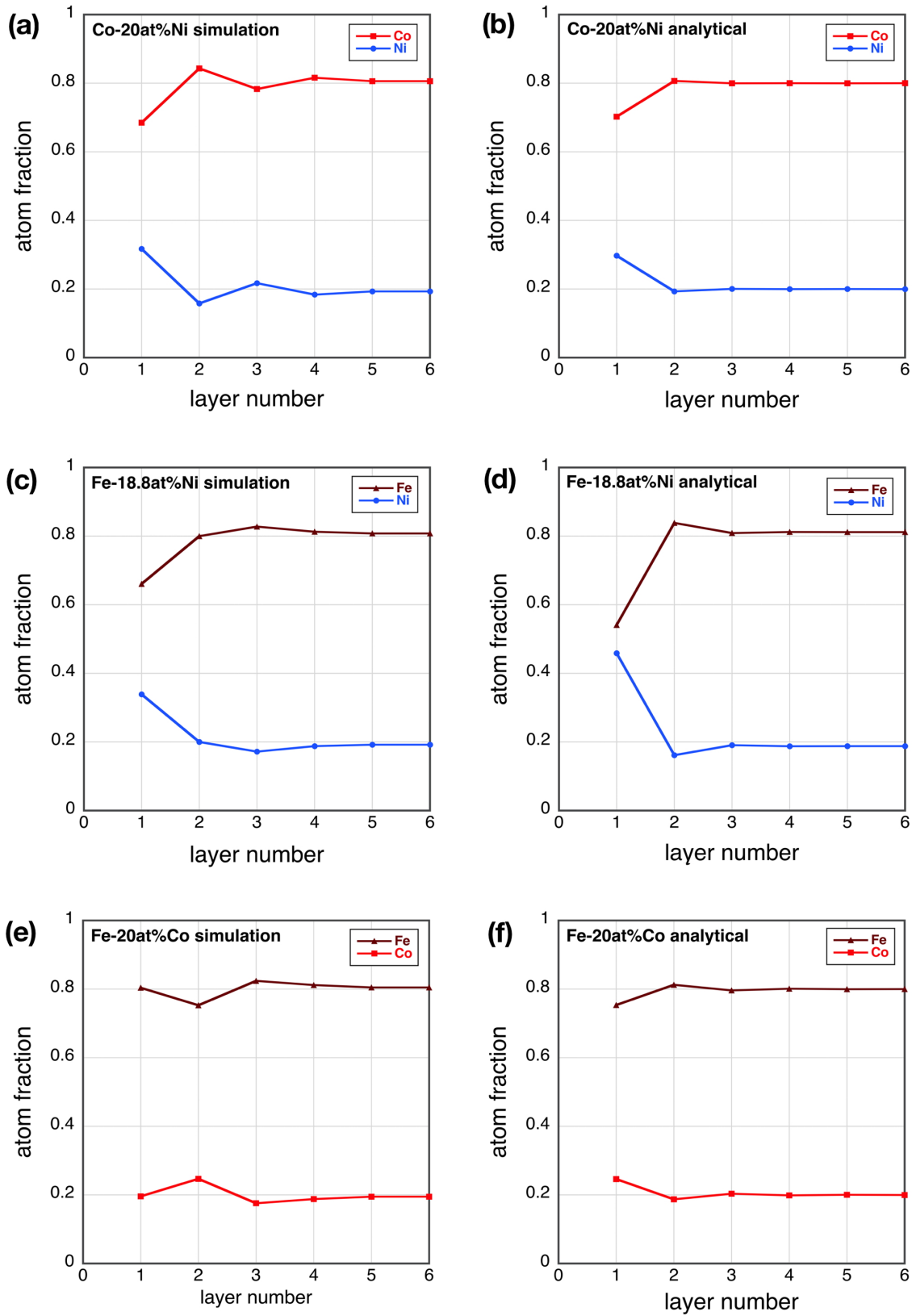

**Fig. 4**. (a) to (d) Near-surface composition profiles of systems with Ni solute, the left column gives the results of simulations, and right column those computed by the analytical model. (e) and (f) Results on the Fe-20 at% Co alloy.

Although no simulations for the Cr-Mn and Fe-Cr binaries could be performed at 1200 K in the FCC structure, due to the instability of FCC, it was possible to obtain segregation profiles for those systems in the FCC structure by means of the analytical model. This is because crystal structure in the analytical model can be fixed arbitrarily by selecting appropriate values of the coordination numbers $z_l$ and $z_v$. Indeed, this is another advantage of the analytical model versus the computer simulations. However, even in the analytical model, in order to obtain physically meaningful results, it is necessary for the temperature and composition to fall within a single-phase domain of the phase diagram. Thus, we have computed the solubility limit of Cr in FCC Fe, at 1200 K, to be slightly above 6 at%, by making use of the regular solution parameter taken from Table 2, and have calculated the surface segregation profile at the 6 at% Cr composition. Results obtained for Cr-20 at% Mn and Fe-6 at% Cr are shown in Figs. 5a and 5b, respectively.

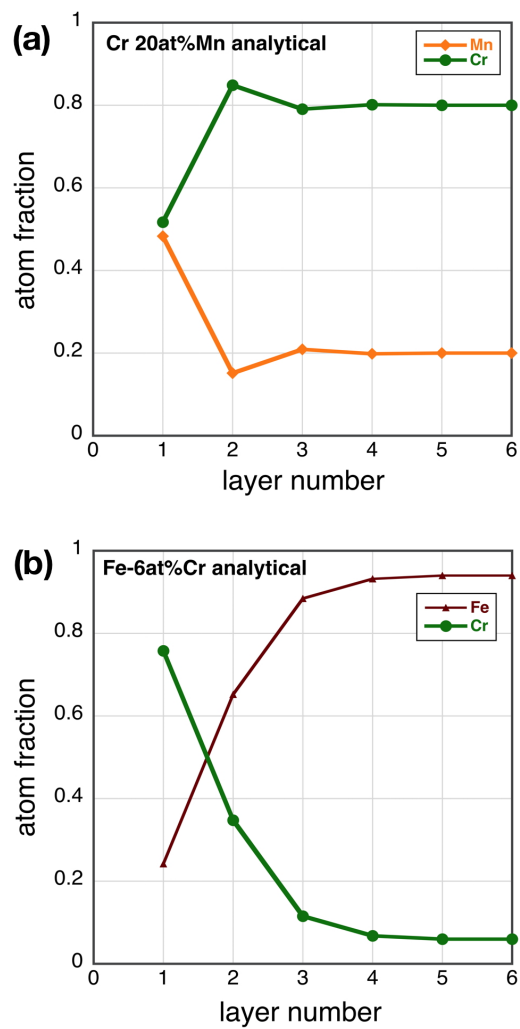

**Fig. 5**. Near-surface composition profiles for (a) Cr-20 at% Mn and (b) Fe-6 at% Cr at 1200 K, calculated in the FCC structure by means of the analytical model.

## 4. DISCUSSION

It is useful to begin by describing the relationship between the composition profiles obtained by the analytical model in the case of the binary alloys, and the input parameters of the model, because that relationship is more transparent than in the case of the profiles obtained by simulations. In binary alloys, surface segregation in the analytical model is driven by three principal terms, the difference in surface energies, the magnitude of regular solution parameter, and the elastic strain energy (see Eq. 2). As mentioned, the latter term has been omitted in the present calculations because of the negligible size difference between all of the five elements of interest [1].

The difference in surface energy between the two components of a binary alloy drives the component of lower surface energy to the surface. For a FCC surface of (001) orientation, where only the outermost surface layer displays nearest-neighbor dangling bonds, the principal effect of the surface energy term is to determine the composition of the outermost surface layer.

The effects of the regular solution parameter differ depending on its sign. A positive parameter corresponds to a clustering type alloy, i.e. an alloy where there is a higher probability for neighbors of an atom to be of the same type. Thus, once atoms of one type, say type-A, have been driven to the outermost surface layer by the surface energy term, then there is a tendency for the next layers to be enriched in type-A atoms. According to Table 2, Fe-Cr is an example of an alloy which displays a strongly positive regular solution parameter. The near-surface composition profile for this alloy, displayed in Fig. 5b, shows that Cr is segregated to the first three atom layers.

A negative regular solution parameter corresponds to an ordering type of alloy, i.e. an alloy where type-A atoms tend to have a preference for type-B neighbors, and vice versa. Thus, if type-A atoms are driven to the outermost surface layer by the surface energy term then the next layer will tend to be enriched in type-B atoms. This effect is displayed, for example, by the Co-Mn alloy in Fig. 2d, where the composition profile is seen to oscillate about the bulk composition. Such oscillation is then evidence of a negative regular solution parameter.

A zero regular solution parameter signifies an ideal solution, in which the probability of a type-A atom having a type-B neighbor is just the atom fraction of type-B atoms in the alloy. As mentioned above, for a FCC (001) surface, in the nearest neighbor bond approximation, only the outermost surface atom layer is affected by the surface energy term, and so one would expect just the outermost atom layer to undergo segregation. An example of that is shown for the Ni-Cr alloy in Fig. 3b, which has a close to zero regular solution parameter, as indicated in Table 2.

Let us now compare the results obtained by simulations to those generated by the analytical model. Figure 1 provides a comparison between the composition profiles obtained by MEAM-based simulations (Fig. 1a) and by the analytical model (Fig. 1b) for the five-component Cantor alloy. One cannot expect complete agreement, because the analytical model consists exclusively of a combination of binary interactions, whereas the MEAM potentials used in the simulations have taken ternary interactions into account, which include the degree of screening by a third element of the interaction between a pair of atoms of two other species [7]. In order to rank the relative strengths of surface segregation of the five elements, we have computed the total adsorption, $\Gamma_n$ of each nth component summed over the L near-surface atom layers:

$$\Gamma_n = \sum_{i=1}^{L}(y_n^i - x_n) \tag{4}$$

The adsorption is the most suitable measure of interfacial segregation, as it is the principal thermodynamic variable of Gibbsian interface thermodynamics [8]. The values of $\Gamma_n$ obtained from Figs. 1a and 1b are given in Table 3.

**Table 3.**
Adsorption of the 5 Cantor alloy components from Figs. 1a and 1b.

| Element | Adsorption (monolayers) from Fig. 1a | Adsorption (monolayers) from Fig. 1b |
|---|---|---|
| Mn | 0.357 | 0.309 |
| Cr | 0.253 | 0.214 |
| Ni | -0.049 | -0.053 |
| Co | -0.197 | -0.161 |
| Fe | -0.353 | -0.308 |

In general, the results of Table 3 indicate that segregation for any given element is relatively weak, ranging over only ±0.35 monolayers. But this result is not too surprising, because in an alloy with five components, all of which are present in relatively high concentration, there is necessarily strong competition for the available surface adsorption sites, as has been observed previously in ternary alloys [22,23]. The most interesting feature of the analytical model is that it not only ranks the segregants in an identical order to that of the simulations, but also gives magnitudes of adsorption that agree remarkably well.

We now turn to a comparison between composition profiles obtained by simulations and by the analytical model, for the case of the underlying binary alloys. The near surface composition profiles are compared in Figs. 2 to 4, and the values of solute adsorption are given in Table 4. Figure 2 compares the alloys with Mn as solute. For those cases, the agreement between the results of simulations and the analytical model is very good, both in the general appearance of the profiles as well as in the respective adsorptions. In all cases Mn exhibits a positive adsorption. This is consistent with the result shown in Fig. 1, that Mn is the strongest segregant in the five-component Cantor alloy. The order of the strength of segregation, with Mn adsorption increasing progressively from Ni-Mn to Co-Mn to Fe-Mn, shown in Table 4, is also consistent with the behavior of the Cantor alloy in Fig. 1.

**Table 4.**
Adsorption for the binary systems from Figs. 2, 3 and 4.

| Binary alloys, figure number | Solute adsorption (monolayers) from simulation | Solute adsorption (monolayers) from analytical model |
| --- | --- | --- |
| Ni-20 at% Mn, Fig. 2 | 0.032 | 0.044 |
| Co-20 at% Mn, Fig. 2 | 0.198 | 0.198 |
| Fe-17.4 at% Mn, Fig. 2 | 0.545 | 0.574 |
| Ni-20 at% Cr, Fig. 3 | -0.030 | -0.092 |
| Co-20 at% Cr, Fig. 3 | 0.297 | 0.099 |
| Co-20 at% Ni, Fig. 4 | 0.104 | 0.091 |
| Fe-18.8 at% Ni, Fig. 4 | 0.131 | 0.200 |
| Fe-20 at% Co, Fig. 4 | 0.027 | 0.036 |

Figure 3 compares the simulation and analytical model results for alloys with a Cr solute, namely Ni-Cr and Co-Cr. Here, there appears to be a systematic discrepancy between the simulation and analytical results. In both alloys, the simulation results display an oscillating composition profile characteristic of a negative regular solution parameter, whereas the analytical model results exhibit either a monotonic variation of the composition profile, as in Co-Cr, or a profile characteristic of an ideal solution with a zero regular solution parameter, as in Ni-Cr. The results of the analytical model are consistent with the values of the regular solution parameters reported in Table 2. However, the adsorption of the Cr solute in Ni-Cr is negative in both the simulation and the analytical model. This is not consistent with the results of Fig. 1 for the quinary alloy, in which Cr segregates more strongly than Ni, and would therefore be expected to have a positive adsorption in a Ni-Cr alloy. The reason for this discrepancy is presumably related to the fact that the MEAM potentials used in the simulations include some ternary interactions that are not explicitly present in the analytical model.

Figure 4 compares the simulation and analytical model results for alloys with a Ni solute, i.e. the Co-Ni and Fe-Ni alloys, and also the single alloy with a Co solute, Fe-Co. Here again, there are some differences in profile shape, but overall the adsorption, which is the principal determinant of thermodynamic behavior, is not significantly different whether it is obtained from the simulations or from the analytical model.

Lastly, it worth commenting on the results obtained with the analytical model for the Cr-Mn and Fe-Cr systems, shown in Figs. 5a and 5b, for which simulations could not be performed in the FCC structure. In Fig. 5a, the Cr-20 at% Mn shows strong segregation of Mn, and Fig. 5b for the Fe-6 at% Cr alloy shows strong segregation of Cr. Both of those results are consistent with the simulations on the quinary in Fig. 1a, in which Mn segregates more strongly than Cr, and where Cr segregates more strongly than Fe. Thus, overall, the results of the analytical model on all the binary systems are consistent with the results of simulations on the quinary, with the exception of the negative adsorption of Cr in the Ni-Cr system.

Given the relative ease of utilizing the analytical model presented here, and the comparable results obtained by that model and the computer simulations, it seems reasonable to use the analytical model as a means of rapidly screening the behavior of complex multi-component systems, so as to identify regions of composition space that might require more detailed examination by massive simulation in order to facilitate surface alloy design. In particular, given that the adsorptions computed by means of the analytical model correspond rather well with those obtained by simulations, the surface energies inferred from the analytical model should also be quite reliable [24,25].

Note that discrepancies between the atomistic and analytical models may arise in cases where strong ternary interactions and/or strong elastic effects are present.

## 5. SUMMARY AND CONCLUSIONS

This paper has compared two approaches for the calculation of the near-surface composition profile that arises from surface segregation in a five-component alloy: atomistic computer simulations in conjunction with MEAM potentials, and an analytical model based on a nearest neighbor bond description, in the regular solution approximation, in which the surface is described as a multilayer region in order to obey the Gibbs adsorption equation. More generally, this model allows the calculation of surface composition profiles and surface energy in a N-component system. Although the analytical approach is presumably less reliable, it can yield results in a few seconds on a laptop computer, whereas the simulation approach requires the use of ~100 parallel processors for 2 to 3 hours on a computation cell consisting of 8000 atoms. The comparison between results obtained by the two approaches was first performed directly on the five-component Cantor alloy. The relative strength of the segregation of the different components was determined from the relative values of Gibbsian adsorption. It was found that not only did the two approaches rank the strength of the segregation of the components in the same order, but the values of the adsorption were also in good agreement. These encouraging results imply that if it is necessary to perform massive simulations of interfacial segregation on multi-component systems, where several different compositions need to be examined in order to perform alloy design, then, an initial screening of the composition space would benefit significantly from a preliminary assessment of the system behavior by means of the analytical model.

The possibility of inferring the behavior of the five-component system from the behavior of the 10 underlying binary systems has also been investigated. The advantage of dealing with binary rather than quinary systems, is that the thermodynamic parameters required for the evaluation of the binaries are far more accessible. Where it was possible to perform the calculations by both approaches, it was found that the adsorptions were in good agreement. However, in the case of segregation in Ni-Cr binary system, it was found that the adsorption of Ni was stronger than that of Cr, whereas the results on the quinary alloy showed a stronger adsorption for Cr. This may be the result of unaccounted ternary element interactions.

Finally, the analytical model offers an approach which is comparable to the use of CALPHAD for calculating phase diagrams; it is useful when the available inputs are limited to macroscopic data (such as surface energy and regular solution parameters). However, the atomistic approach remains essential when it becomes necessary to "observe" the atomic structure in detail, such as for example to detect the presence of dislocations and/or other defects that may be present at segregated interfaces.

## ACKNOWLEDGMENTS

PW acknowledges with thanks use of the resources of the National Energy Research Scientific Computing Center (NERSC), a US Department of Energy Office of Science User Facility operated under Contract No. DE-AC02-05CH11231. DC wishes to acknowledge the French Agence Nationale de la Recherche for support of her research under grant ANR-AHEAD-16-CE92-0015-01 in the framework of the ANR-DFG project: Analysis of the Stability of High Entropy Alloys by Dewetting of Thin Films (AHEAD). Both authors also wish to thank Dr. Céline Varvenne (CINaM), for fruitful discussions.

## DATA AVAILABILITY STATEMENT

This paper does not use any raw or processed data. It strictly reports calculations. It includes all the data necessary to perform the calculations and support the findings of this study.